\documentclass[reprint,aps,twocolumns,superscriptaddress,floatfix,notitlepage]{revtex4-1}
\usepackage{feynmp-auto}

\usepackage[a4paper]{geometry}
\usepackage{graphicx}
\usepackage{float}
\usepackage{amsmath}
\usepackage{amsfonts}
\usepackage{amssymb}
\usepackage{enumitem}
\usepackage{epstopdf}
\usepackage{mathrsfs}
\usepackage{mathtools}
\usepackage{fullpage}
\usepackage[utf8]{inputenc}
\usepackage{xcolor}
\usepackage{bbold}
\usepackage{physics}
\usepackage{bm}

\def\>{\rangle}
\def\<{\langle}

\newcommand\+{\dagger}
\newcommand\x{\mathbf{x}}

\newcommand\mk{\mathbf{k}}

\newcommand\mq{\mathbf{q}}

\newcommand\vp{\varphi}

\newcommand{\be}{\begin{equation}}
\newcommand{\ee}{\end{equation}}
\newcommand{\bea}{\begin{eqnarray}}
\newcommand{\eea}{\end{eqnarray}}
\newcommand{\p}{\partial}

\renewcommand{\vec}[1]{\mathbf{#1}}
\newcommand*\diff{\mathop{}\!\mathrm{d}}
\usepackage{braket}
\usepackage[colorlinks,bookmarks=true,citecolor=blue,linkcolor=red,urlcolor=blue]{hyperref}
\usepackage{hyperref}

\begin{document}

    \title{Transport properties of multilayer graphene}
	\author{Glenn Wagner}
	\affiliation{Rudolf Peierls Centre for Theoretical Physics, Parks Road, Oxford, OX1 3PU, UK}
	\author{Dung X. Nguyen}
    \affiliation{Brown Theoretical Physics Center and Department of Physics,
Brown University, 182 Hope Street, Providence, RI 02912, USA}
	\author{Steven H. Simon}
	\affiliation{Rudolf Peierls Centre for Theoretical Physics, Parks Road, Oxford, OX1 3PU, UK}

	\begin{abstract}
    We apply the semi-classical quantum Boltzmann formalism for the computation of transport properties to multilayer graphene. We compute the electrical conductivity as well as the thermal conductivity and thermopower for Bernal-stacked multilayers with an even number of layers. We show that the window for hydrodynamic transport in multilayer graphene is similar to the case of bilayer graphene. We introduce a simple hydrodynamic model which we dub the multi-fluid model and which can be used to reproduce the results for the electrical conductivity and thermopower from the quantum Boltzmann equation.
	\end{abstract}
	\maketitle	

\section{Introduction}

Ultra-clean materials such as graphene offer a new perspective on electronic transport. At low enough temperatures, momentum-relaxing scattering of the electrons such as the scattering off phonons or impurities is sub-dominant and the dominant source of collisions are the collisions of the electrons with themselves. This is the realm of electron hydrodynamics \cite{Lucas_2018}. In the hydrodynamic regime the electron-electron scattering rate $\tau_{ee}^{-1}$ is larger than the electron-phonon scattering rate $\tau_{ep}^{-1}$. Both monolayer and bilayer graphene have garnered much attention in recent years for their supposed hydrodynamic transport \cite{Bandurin1055,Levitov2016,Sulpizio2019,Svintsov,Pellegrino2017,Hydro1,Hydro2,Hydro3,Hydro4}. In the present paper, we focus on a related material: multilayer graphene. 

The Boltzmann equation is an equation of motion for the distribution function of particles and has traditionally been applied to the classical kinetic theory of gases. Extending this approach to the study of electron gases in graphene and related materials leads to the celebrated quantum Boltzmann equation (QBE) \cite{Greenwood1958,Sachdev,Lux2013,Lux2012,Dumitrescu:2015,Mueller2008b,Vignale2,Zarenia_2019}. 

In recent work, the present authors presented the QBE formalism for bilayer graphene \cite{ourPRB}. It was then shown in Ref.~\cite{ourPRL} that the QBE results agree well with experimental measurements of the electrical conductivity of suspended bilayer graphene in \cite{Nam2017}. Despite its success, the QBE is a heavy-handed approach and this led to the development of the two-fluid model. In bilayer graphene, the low-energy bandstructure consists of two gapless quadratic bands which can be populated with electrons and holes. The dynamics of the electron and hole fluids can be captured accurately from simple hydrodynamic equations, at least for the calculation of the electrical conductivity and thermopower \cite{ourPRL}.

In this work, we generalize the formalism we developed for BLG to multilayer graphene (MLG), in particular, we focus on Bernal (AB) stacked multilayers. We consider the special case of an even number of layers $N$, to avoid the additional complication of the linear band that arises for odd $N$.  We study the regime near charge neutrality, i.e. $\beta\mu\lesssim1$, where $\beta$ is inverse temperature and $\mu$ is the chemical potential. In fact, in this regime, we expect the behaviour for even $N$ and odd $N+1$ to be very similar, since the density of states is dominated by the quadratic bands. We use the QBE to compute the electrical conductivity, the thermal conductivity and thermopower for multilayers with $N=2$ to $N=8$ layers. We discuss how the transport properties evolve, as the number $N$ of layers is increased. In particular, in previous work \cite{ourPRL,ourPRB}, the present authors discussed two signatures of the hydrodynamic regime: The Wiedemann-Franz law violation and the fast increase of the electrical conductivity away from charge neutrality. We will show that both of the signatures remain, as we increase the number of layers in our graphene multilayer. We then develop a hydrodynamic approach in terms of a multicomponent fluid and show that it accurately matches the QBE predictions.

There has been previous theoretical work on transport in multilayer graphene using the Kubo formula \cite{zeroT}. Ref.~\cite{imp} does study MLG using the Boltzmann formalism, however, they focus on the case where impurities are the main source of scattering. Ref.~\cite{phonons} calculates the thermal conductivity due to phonons, however they do not explore the electronic contribution to the thermal conductivity.

The electrical conductivity of multilayer graphene has been measured experimentally for a range of temperatures and densities \cite{Nam_2016,Nam324}. In Ref.~\cite{wang}, measurements on the minimum of the electrical conductivity for different numbers of layers are reported. Further experiments by a different experimental group have been reported \cite{Ye2011}, however, they consider the high-density regime which is the opposite limit to the one we will consider in this work.

The structure of our paper is as follows. We start by introducing the tight-binding model for MLG and we discuss the screening of the Coulomb interaction in MLG. We then introduce the QBE formalism and present the numerical results for different numbers of layers. Finally, we discuss how many of the salient features of our numerics can be captured by a simple hydrodynamic model.

\section{Bandstructure and interactions}
\begin{figure}
    \centering
    \includegraphics[width=0.4\textwidth]{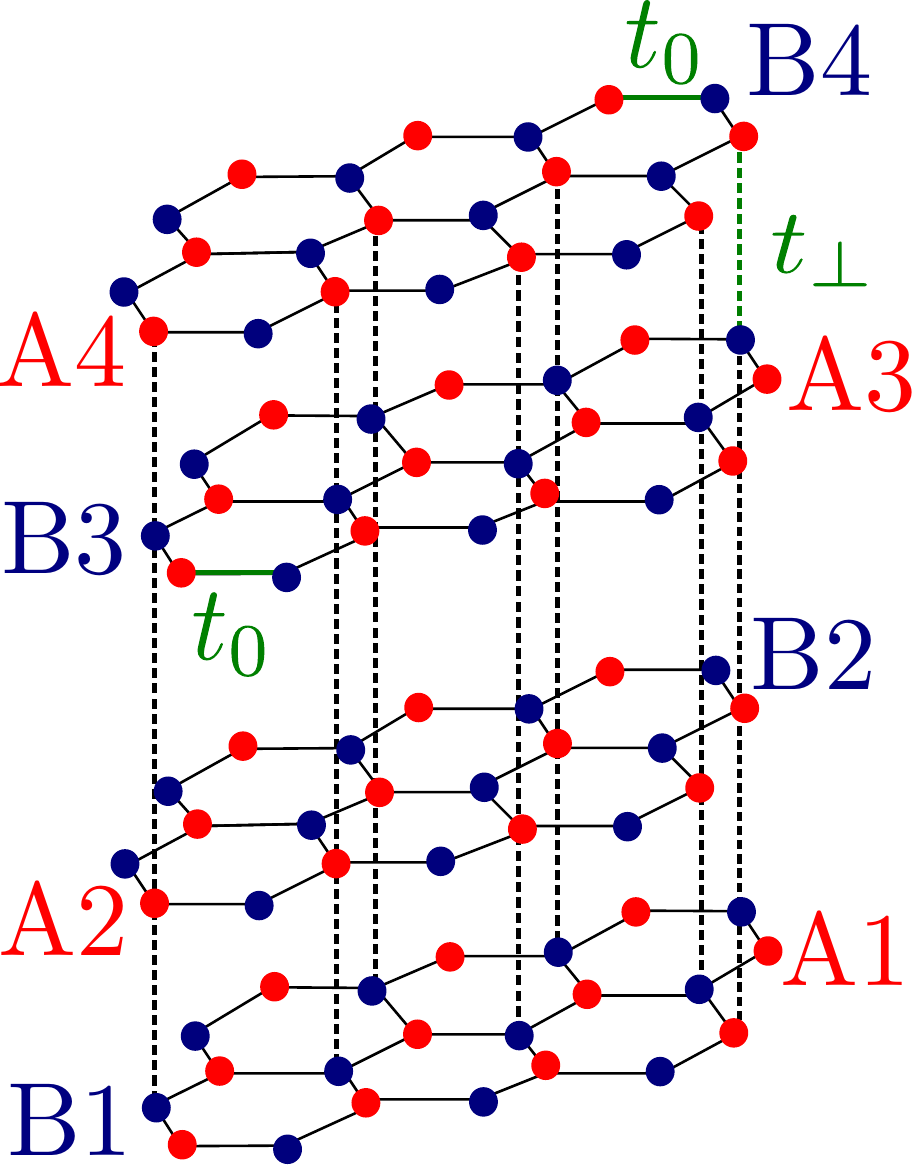}
    \caption{AB stacked multilayer with $N$=4 layers. We use a tight-binding model with nearest neighbour intralayer ($t_0$) and interlayer ($t_\perp$) hopping.}
    \label{fig:model}
\end{figure}

For the Bernal (AB) stacking of $N$ graphene multilayers shown in Fig.~\ref{fig:model}, the tight-binding (Bloch) Hamiltonian expanded near the $K$ and $K'$ valleys is the $2N\times 2N$ matrix \cite{Min2008,zeroT,Koshino2007,Latil,Guinea2006,Ohta,Nilsson,Koshino2010} 

\begin{equation*}
   H= \left(
\begin{array}{ccccccccc}
 0 & v\pi^\dagger & 0 & 0 & 0 & 0 & 0 & 0 & \\
 v\pi & 0 & t_\perp  & 0 & 0 & 0 & 0 & 0 & \\
 0 & t_\perp  & 0 & v\pi^\dagger & 0 & t_\perp  & 0 & 0& \\
 0 & 0 & v\pi & 0 & 0 & 0 & 0 & 0& \\
 0 & 0 & 0 & 0 & 0 & v\pi^\dagger& 0 & 0& \\
 0 & 0 & t_\perp  & 0 & v\pi & 0 & t_\perp  & 0 & \\
 0 & 0 & 0 & 0 & 0 & t_\perp  & 0 & v\pi^\dagger&  \\
 0 & 0 & 0 & 0 & 0 & 0 & v\pi & 0& \\
  &  &  &  &  &  &  & & \ddots \\
\end{array}
\right).
\end{equation*}
Here $\pi=\xi p_x+i p_y$, with valley index $\xi=\pm$. The Fermi velocity is $v=\frac{\sqrt{3}}{2}at_0$, where $t_0$ is the intralayer hopping parameter and $a$ is the lattice constant. $t_\perp$ is the interlayer hopping. The wavefunction is $\psi=(\vp_{A_1},\vp_{B_1},\vp_{A_2},\vp_{B_2},\cdots,\vp_{A_N},\vp_{B_N})$, where $\vp_{A_i(B_i)}$ is the wave function of an electron at site $A_i(B_i)$.

At low energies, we can focus on the gapless bands. We write down a low energy $N\times N$ Hamiltonian for these $N$ quadratic bands, which we label with $r=(R,\sigma)$ where $R=1,\dots, N/2$ and $\sigma=\pm 1$. The energies are
\begin{equation}
    \varepsilon_{R\sigma}(p)=\sigma \frac{p^2}{2m_R},
\end{equation}
where the mass is 
\begin{equation}
    m_R=2m^* \ \textrm{cos}\left(\frac{R\pi}{N+1}\right),
\end{equation}
with $m^*=t_\perp/2v^2$. Therefore the bands appear in pairs labelled by the same $R$ which have the same mass. For a fixed $R$, we have the same bandstructure as in BLG. We call the corresponding Bloch wavefunctions $|\psi_{R\sigma}(\vec p)\rangle$. The matrix elements for the Bloch functions are
\begin{align}
\label{eq:M}
    M_{R\sigma,R'\sigma'}(\vec p,\vec p')&\equiv\langle \psi_{R'\sigma'}(\vec p')|\psi_{R\sigma}(\vec p)\rangle\\&=\frac{\delta_{RR'}}{2}(1+\sigma\sigma'e^{-2i(\theta_p-\theta_{p'})}),\nonumber
\end{align}
where $\cos\theta_p=\vec p\cdot\hat{\vec{x}}$. The derivations of the effective mass $m_R$ as well as the matrix elements \eqref{eq:M} are left for Appendix \ref{sec:TB}. There will be no vertex coupling electrons with different $R$ in the Coulomb interaction. For a pair of bands with the same value of $R$, the matrix elements are the same as for BLG. Using this result, one can perform the classic Lindhart calculation for the polarization $\Pi^0(\Vec{q},\omega)$ in the limit $\beta\mu\lesssim1$ and $\beta q^2/m\lesssim 1$. This is a calculation analogous to Refs.~\cite{Lv,Min2012} and the details are in Appendix \ref{sec:Screen}. We focus on the static polarization, which is valid at low enough temperatures. The result for the polarization is
\begin{equation}
    \Pi^0(\Vec{q},0)=-\frac{N_fm^*}{2\pi}\bigg(\frac{1}{\sin(\frac{\pi}{2(N+1)})}-1\bigg),
\end{equation}
where $N_f=2\times 2$ accounts for the spin and valley degrees of freedom. In the screening calculation we have assumed that the screening due to the phonons is negligible. The Thomas-Fermi screening wavevector is then given by 
\begin{equation}
q_{TF}(q)=-\Pi^0(\mq,0) 2\pi \alpha,
\end{equation}
where $\alpha$ is the electromagnetic fine-structure constant. We will use the fully screened Coulomb interaction $V(q)=2\pi \alpha/q_{TF}(q)$, which is a good approximation at low temperatures, where the typical momentum of electrons is much smaller than $q_{TF}$. Now we approximately have the behaviour $q_{TF}\propto N$. Since each electron can now scatter off $N$ species of electrons, the electron-electron scattering rate will be $\tau_{ee}^{-1}\propto N/q_{TF}^2\propto 1/N$. 

\section{Quantum Boltzmann equation}

Away from charge neutrality, one needs to include momentum-relaxing scattering in order to obtain a well-defined conductivity. Based on the results in bilayer graphene, we expect electron-phonon collisions to be the dominant source of momentum-relaxing scattering and hence this is the only momentum-relaxing mechanism that we include in our calculations \cite{ourPRL}. Depending on the experimental conditions, we may envisage electron-impurity and electron-boundary scattering as well and this would be a simple extension of the present calculation. The phonon scattering is proportional to band mass and we extract the proportionality constant by comparing the BLG results to available experimental data \cite{ourPRL}. The QBE is an evolution equation for the distribution function $f_{r}(\mk,\x,t)$ of the particles of species $r$ of the form
	\begin{multline}
	\label{eq:Boltz1}
	\left(\frac{\p}{\p t}+\mathbf{v}_r(\mathbf{k}) \cdot\frac{\p}{\p \x} +e\mathbf{E}  \cdot \frac{\p}{\p \mk}\right)f_{r}(\mk,\x,t)\\=-I_{r}[\{f_{r_i}\}](\mk,\mathbf{x},t),
	\end{multline}
where $\mathbf{v}_{r}(\mathbf{k})=\p_{\mathbf{k}}\epsilon_{r}(\mathbf{k})$, $e<0$ is the electron charge and the collision integral on the RHS includes electron-electron and electron-phonon collisions. The electron-electron collision integral can be derived from the Kadanoff-Baym equations \cite{Kadanoff} using the Born approximation. The derivation is identical to the BLG case in Ref.~\cite{ourPRB}. The electron-phonon collision integral uses the simple relaxation-time approximation with scattering rate 
    \begin{equation}
        \tau_{ep,r}^{-1}=\frac{D^2m_r k_BT}{2\rho\hbar^3c^2},
    \end{equation}
where $D$ is the deformation potential, $\rho$ is the mass density of multilayer graphene and $c$ is the speed of sound. We also define the corresponding dimensionless parameter $\alpha_{ep}=\beta\tau_{ep}=\beta m^*/m_r\tau_{ep,r}^{-1}$. The full details of the QBE are shown in appendix \ref{sec:QBE}. We note that $\rho\propto N$ and $c=\textrm{const.}$, so assuming that $D$ only depends weakly on $N$, we have $\alpha_{ep}\propto 1/N$. We now see that both the electron-electron and the electron-phonon scattering rates behave like $1/N$, although the reasons behind this scaling are very different for the two scattering mechanisms. Based on this simple scaling, it stands to reason that the hydrodynamic window in multilayer graphene is similar to that of BLG: $\tau_{ee}^{-1}/\tau_{ep}^{-1}$ is only weakly $N$-dependent. Since we successfully applied a hydrodynamic model to BLG \cite{ourPRL}, we expect this to work for MLG as well.

\begin{figure}
    \centering
    \includegraphics[width=\linewidth]{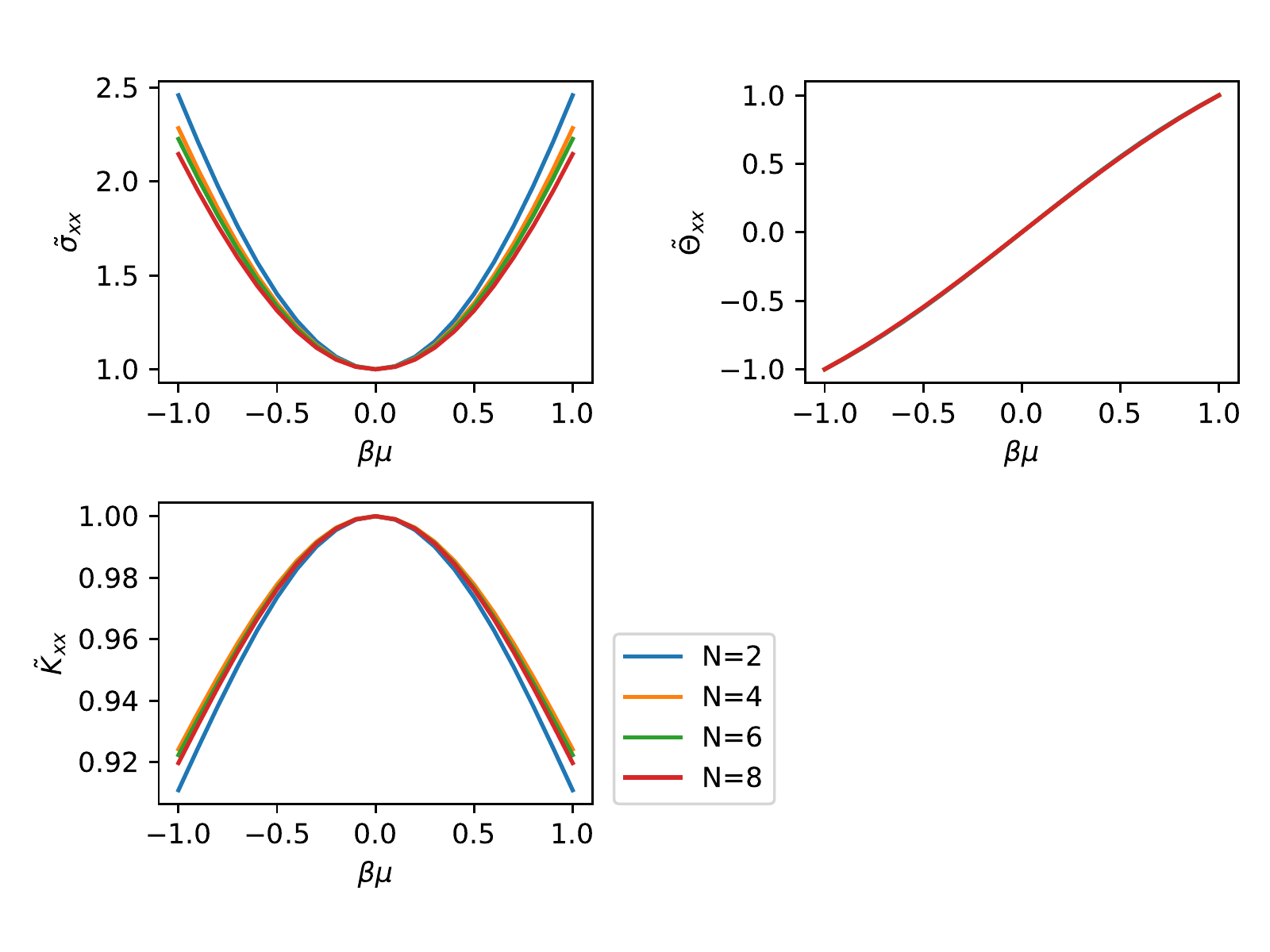}
    \caption{Results from the QBE calculation for different even values of $N$. We plot the normalized electrical conductivity $\tilde\sigma_{xx}(\beta\mu)/\tilde\sigma_{xx}(\beta\mu=0)$, thermal conductivity $\tilde K_{xx}(\beta\mu)/\tilde K_{xx}(\beta\mu=0)$ and thermopower $\tilde\Theta_{xx}(\beta\mu)/\tilde\Theta_{xx}(\beta\mu=1)$. We have set $\alpha_{ep}=0.1/N$ as in our previous work on BLG. }
    \label{fig:QBE_norm}
\end{figure}

In order to solve the QBE, we expand the distribution function in terms of $4N$ basis functions. Based on our previous work \cite{ourPRL} this is a sufficient number of basis functions to obtain a convergent result. The QBE then turns into an equation for the expansion coefficients in front of the basis functions.
Once we know the perturbation of the distribution function due to an applied thermal gradient $\nabla T$ or electric field $\vec E$, we can compute the electrical current 
		\begin{equation}
	\label{eq:currentexpect}
	  \mathbf{J}=  N_fe\sum_{r}\int \frac{d^2k}{(2\pi)^2} \frac{\sigma \mk}{m_r} f_{r}(\mk),
	\end{equation}
	and heat current
		\begin{equation}
	\label{eq:Qcurrentexpect}
	  \mathbf{J}^{Q}=  N_f\sum_{r}\int \frac{d^2k}{(2\pi)^2} \frac{\sigma \mk}{m_r}(\epsilon_{r}(\mk)-\mu) f_{r}(\mk).
	\end{equation}
We define the electrical conductivity $\sigma$, the thermal conductivity $K$ and the thermopower $\Theta$ by
	\begin{equation}
	\begin{pmatrix}
	\vec{J}\\
	\vec{J}^Q
	\end{pmatrix}
	=\begin{pmatrix}
	\sigma &\Theta\\
	T\Theta&K
	\end{pmatrix}
	\begin{pmatrix}
	\vec{E}\\
	-\vec{\nabla}T
	\end{pmatrix}.
	\label{eq:def_thermo}
	\end{equation}
Note that the open circuit thermal conductivity $\kappa$ measuring the heat current in the absence of electrical current is given by $\kappa=K-T\Theta\sigma^{-1}\Theta$. In the absence of a magnetic field, the transport coefficients are diagonal. In Fig.~\ref{fig:QBE_norm} we plot the dimensionless transport coefficients
\begin{equation}
    \tilde\sigma_{xx}\equiv \frac{2}{N_fe^2} \sigma_{xx},
\end{equation}
\begin{equation}
    \tilde\Theta_{xx}\equiv \frac{2}{N_fek_B} \Theta_{xx},
\end{equation}
\begin{equation}
    \tilde K_{xx}\equiv \frac{2}{N_fk_B^2T} K_{xx}
\end{equation}
for different values of $N$. 
\begin{figure}
	\centering
	\includegraphics[width=\linewidth]{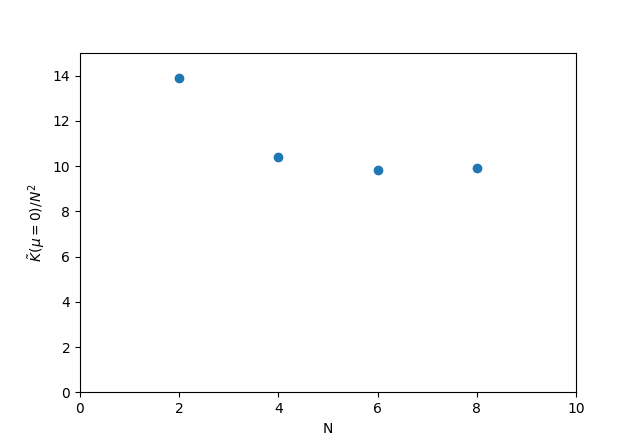}
	\caption{Results for the thermal conductivity $\tilde K(\mu=0)/N^2$ from the QBE calculation for different even values of $N$. We have set $\alpha_{ep}=0.1/N$ as in our previous work on BLG.}
	\label{fig:K_N}
\end{figure}

If we could treat the $N$-layer multilayer as $N/2$ independent bilayers, then we would expect the transport coefficients to increase proportionally to $N$. However, this is not what happens. Indeed we find approximately $K(\beta\mu=0)\propto N^2$ in Fig.~\ref{fig:K_N}. The reason for this behaviour is that $K$ at charge neutrality is limited by collisions with phonons -- it would diverge in the absence of phonon scattering since Coulomb scattering does not relax the mode where all carriers move at the same velocity. Recall the formula from basic kinetic theory $K\sim\Lambda k_B^2 n\tau_{ep} T/m$, where $\Lambda k_B$ is the heat capacity per particle and $n$ is the number density. Now $\tau\propto 1/N$ as explained in the previous section. We also have $n\propto N$. This explains the observed $K(\beta\mu=0)\propto N^2$ behaviour.

\begin{figure}
    \centering
    \includegraphics[width=\linewidth]{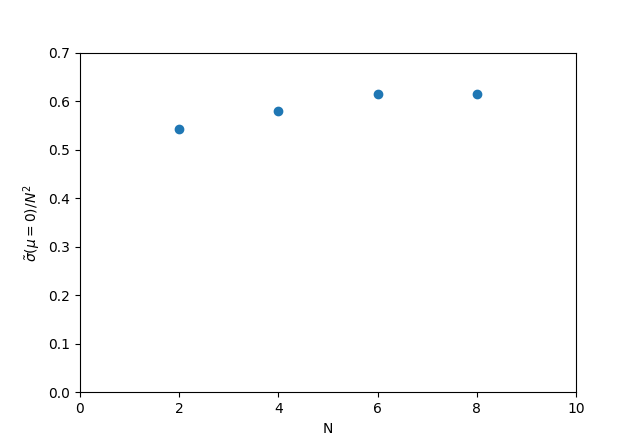}
    \caption{Results for the electrical conductvity $\tilde \sigma(\mu=0)/N^2$ from the QBE calculation for different even values of $N$. We have set $\alpha_{ep}=0.1/N$ as in our previous work on BLG.}
    \label{fig:sigma_N}
\end{figure}

We plot the results for the electrical conductivity at charge neutrality as a function of $N$ in Fig.~\ref{fig:sigma_N}. We observe that the conductivity scales with $N$ approximately as $\sigma(\beta\mu=0)\propto N^2$. Recall the Drude formula $\sigma\sim e^2 n\tau/m$, where $\tau$ is the collision time for current-relaxing collisions. For different species the individual conductivities will add up, ie $\sigma\sim \sum_re^2 n_r\tau_r/m_r$. $\sigma(\beta\mu=0)$ is well-defined even in the absence of phonons and indeed electron-electron collisions will dominate the current relaxation. As we increase $N$, we increase the density of states and hence the screening wavevector scales approximately as $q_{TF}\propto N$ and hence the potential scales as $V\sim 1/N$. Therefore Fermi's Golden rule for scattering of particles of species $r$ off particles of species $r'$ yields $\tau_{rr'}^{-1}\propto |V|^2\propto 1/N^2$. The scattering time for species $r$ then roughly scales as $\tau_r^{-1}\equiv\sum_{r'}\tau_{rr'}^{-1}\propto 1/N$. So $\tau_r\propto N$ and with $n_r\propto m_r$, we find $\sigma\sim \sum_re^2 n_r\tau_r/m_r\propto \sum_r\tau_r\propto N^2$, as in the numerical results.

Let us discuss two signatures of the hydrodynamic regime: (i) the ratio $\sigma(\beta\mu=1)/\sigma(\beta\mu=0)$ and (ii) the Wiedemann-Franz law violation. The ratio $\sigma(1)/\sigma(0)$ stays relatively constant as $N$ is increased. Recall that for bilayer graphene, the reason for the large value of $\sigma(1)/\sigma(0)$ is that $\sigma(\beta\mu=0)$ is limited by electron-electron collisions, which operate on a time-scale $\tau_{ee}$. On the other hand, away from CN the momentum mode carries charge and this momentum mode is relaxed on a much longer time-scale $\tau_{ep}$. Since $\tau_{ep}\gg\tau_{ee}$ in the hydrodynamic regime, charge transport is greatly enhanced away from CN. Since both $\tau_{ee}$ and and $\tau_{ep}$ scale proportional to $N$, $\sigma(1)/\sigma(0)$ does not vary significantly with $N$. In the hydrodynamic regime, the Lorenz number is much larger than predicted by the Wiedemann-Franz law. Since $\sigma$ increases as fast as $K$ with $N$, the violation of the Wiedemann-Franz law at charge neutrality will also remain relatively constant as a function of $N$. We show plots of $\sigma(1)/\sigma(0)$ and the Lorenz number in Appendix \ref{sec:sup_figs}.

In Fig.~\ref{fig:8} we show the results for the three transport properties considered for the representative case of $N=8$. Other even values of $N$ yield similar results. As $N\to\infty$ and we approach the graphite limit, our numerics become unmanageable and a full 3d theory becomes necessary, where the bands are dispersive along $k_z$, instead of having large number $N$ of 2d bands as in our 2d model. In fact, due to the approximations we have made, the low energy theory we have derived is valid in the limit $N\ll 100$ \footnote{In \eqref{eq:eigenenergies}, the gap $\Delta_r=\varepsilon_r^+(p=0)-\varepsilon_r^-(p=0)=2\gamma\cos(r\pi/(N+1))$. The smallest gap therefore occurs when $r=\frac{N/2}{N+1}$ and is $\Delta_\textrm{min}\approx2\gamma\sin(\pi/N)\approx 2\gamma\pi/N$. We want the gap to be so large that we can neglect the gapped bands, i.e. $\Delta_\textrm{min}\gg \frac{1}{\beta}$. Using $\gamma\approx0.3$eV \cite{McCann_2013} and for $T=40K$ this leads to $N\ll 100$. }.

\section{multi-fluid model}

Following the usual procedure for deriving hydrodynamic equations from kinetic theory, we can obtain the fluid equations from the full QBE. We have $r$ species of fermions in the low energy theory, and in the hydrodynamic description, we can associate each fluid species with a mean velocity $\vec u^r$. The equation of motion that follows from the QBE for $\vec u^r$ under an applied electric field $\vec E$ and thermal gradient $\vec \nabla T$ is 
\begin{multline}
    m_r\partial_t\vec{u}^r=-\sum_{r'}\frac{m_r}{\tau_{rr'}}(\vec{u}^r-\vec{u}^{r'})-\frac{m_r\vec{u}^r}{\tau_{ep,r}}\\+\sigma e\vec{E}-\Lambda^r k_B\nabla T,
\end{multline}
where $\tau_{rr'}$ is the effective scattering time of particles of species $r$ off particles of species $r'$ due to Coulomb interactions, $\tau_{ep,r}$ is the effective electron-phonon scattering time for species $r$ and $\Lambda^r$ is the entropy per particle:
\begin{widetext}
\begin{equation} 
\Lambda^{(R,+)}=-\frac{\int \frac{d^{2}\vec p}{(2 \pi)^{2}}\left[\left(1-f^0_r(\mathbf{p})\right) \ln \left(1-f^0_r(\mathbf{p})\right)+f^0_r(\mathbf{p}) \ln f^0_r(\mathbf{p}) \right]}{\int \frac{d^{2} \vec p}{(2 \pi)^{2}} f^0_r(\mathbf{p})} ,
\end{equation}

\begin{equation} 
\Lambda^{(R,-)}=-\frac{\int \frac{d^{2}\vec p}{(2 \pi)^{2}}\left[\left(1-f^0_r(\mathbf{p})\right) \ln \left(1-f^0_r(\mathbf{p})\right)+f^0_r(\mathbf{p}) \ln f^0_r(\mathbf{p}) \right]}{\int \frac{d^{2} \vec p}{(2 \pi)^{2}} [1-f^0_r(\mathbf{p})]}.
\end{equation}
\end{widetext}
Solving the fluid equations for a steady state flow yields an expression for $\vec u^r$. The electrical current is then given by
\begin{equation}
    \vec{J}=e\sum_r\sigma n^r\vec{u}^r
    \label{eq:J_current1}
\end{equation}
and the heat current by
\begin{equation}
    \vec{Q}= k_BT\sum_r\Lambda^rn^r\vec{u}^r.
    \label{eq:Q_current1}
\end{equation}
The detailed derivation is in Appendix \ref{sec:mFM}.

\begin{figure}
    \centering
    \includegraphics[width=\linewidth]{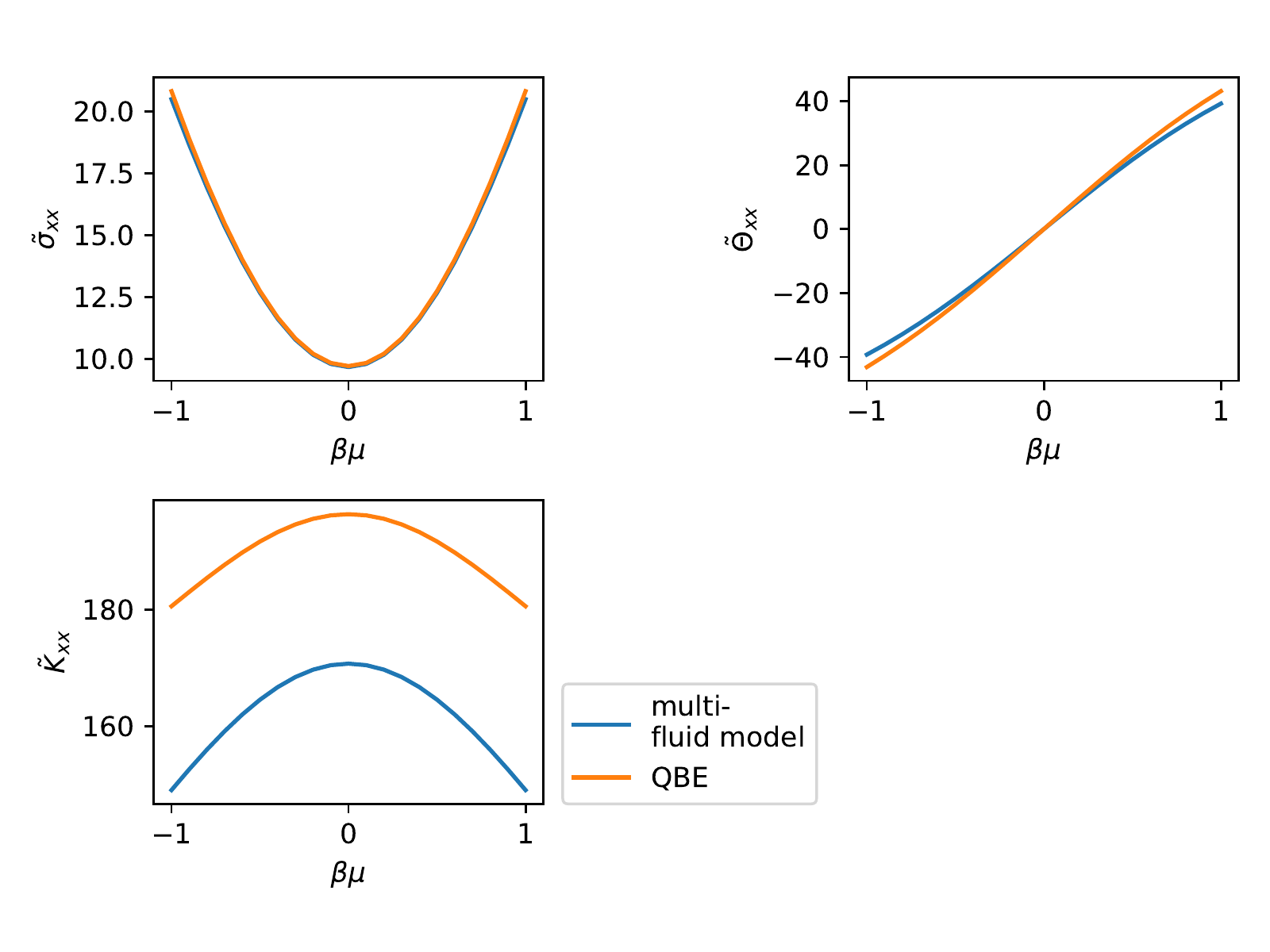}
    \caption{Dimensionless transport coefficient calculated for the representative case $N=8$. Comparison of the QBE results with the multi-fluid model introduced in the main text. The value for the Coulomb scattering strength $\alpha_0$ to use in the multi-fluid model has been extracted from the QBE results, such that $\tilde\sigma_{xx}(\beta\mu=0)$ is the same for both plots (ie we have one fitting parameter). The multi-fluid model performs well for the electrical conductivity and the thermopower, but a bit less well for the thermal conductivity.}
    \label{fig:8}
\end{figure}

In Fig.~\ref{fig:8} we compare the results from the multi-fluid model and the QBE and find that for the electrical conductivity the agreement is excellent, whereas for the thermal conductivity, the qualitative behaviour is correct but the quantitative agreement is off by around 20\%. The reason for this is that the multi-fluid model is equivalent to solving the QBE by using only $N$ basis functions in the expansion of the distribution function. These basis functions correspond to uniform motion with velocity $\vec u^r$ of the fermions of species $r$. For an applied electric field, these modes capture the charge transport accurately, as exemplified by the good overlap in Fig.~\ref{fig:8}. On the other hand, for an applied thermal gradient, we do not accurately capture the heat transport with those modes. We found the same situation in Ref.~\cite{ourPRL} for BLG.

The success of the multi-fluid model as well as the two fluid model in our previous work on BLG \cite{ourPRL} suggests that the hydrodynamic description of electrons in bilayer and multi-layer graphene is accurate. This once again confirms the idea that electrons in strongly interacting systems can be considered as (multi-component) fluids \cite{Lucas_2018}. 

\section{summary}
We have applied the quantum Boltzmann formalism to study the transport properties of multilayer graphene. We find results very similar to bilayer graphene. We introduce a hydrodynamic model which agrees accurately with the QBE results for the electrical conductivity and thermopower. We hope that future experiments on transport in multilayer graphene will reveal whether the QBE formalism performs as well for multilayer graphene as it does for BLG, although we see no apparent reason why it should not. 

We have only studied even $N$ in this paper. For odd $N$, the low energy theory consists of $N-1$ parabolic bands and one Dirac cone. However, in the regime $\beta\mu\lesssim1$ the density of states will be dominated by the quadratic bands. Therefore the results for odd $N$ are expected to be similar to the results for even $N$, as long as one accounts for the different values of the band masses.

The behaviour of the transport properties as the number $N$ of layers is varied shows some interesting features. Firstly, the thermal conductivity at charge neutrality (CN) $K(\beta\mu=0)$ is approximately proportional to $N^2$. This is due to the fact that the thermal conductivity at CN is limited by phonons and the phonon scattering time is proportional to $N$, so $K(\beta\mu=0)\propto n\tau\propto N^2$. The electrical conductivity at CN $\sigma(\beta\mu=0)\propto N^2$ as well, but for a different reason. In contrast to $K(\beta\mu=0)$, $\sigma(\beta\mu=0)$ is limited by electron-electron collisions. As $N$ is increased, the screening increases and so the electron scattering time $\tau\propto N$, leading to $\sigma(\beta\mu=0)\propto n\tau\propto N^2$. Put together, this implies that the violation of the Wiedemann-Franz law stays constant as $N$ is increased. Finally, $\sigma(\beta\mu=1)/\sigma(\beta\mu=0)$, which is another measure of the relative size of the electron-electron and the electron-phonon scattering times. is relatively flat as a function of $N$.

In future work, we plan to compute the viscosity for MLG. Adding the viscosity to the multi-fluid model will give us the Navier-Stokes equations, which can then be used to simulate the electron fluid in MLG for realistic geometries. We expect those simulations to yield interesting results such as the vortices which have been predicted for single-layer graphene \cite{Levitov2016} and negative resistivity, which has been seen in experiments in single-layer graphene \cite{Bandurin1055}. One can go even further and consider spin-transport by applying a weak magnetic field. We then have a very interesting  multi-component fluid which carries charge, heat and spin.
\\
\\
\section{Acknowledgements}

This work was supported by EPSRC grants EP/N01930X/1 and EP/S020527/1. DXN was supported partially by Brown Theoretical Physics Center.

\appendix

\begin{widetext}

\section{Derivation of effective mass $m_r$ and matrix element $M_{r r'}(\mathbf{p},\mathbf{p}')$}
\label{sec:TB}

\subsection{Low-energy band theory and matrix elements}
We use the effective Hamiltonian from Ref.~\cite{Min2008}. For Bernal (AB) stacking of $N$ graphene multilayers, the tight-binding Hamiltonian is the $2N\times 2N$ matrix

\begin{equation}
   H= \left(
\begin{array}{ccccccccc}
 0 & v\pi^\dagger & 0 & 0 & 0 & 0 & 0 & 0 & \\
 v\pi & 0 & \gamma  & 0 & 0 & 0 & 0 & 0 & \\
 0 & \gamma  & 0 & v\pi^\dagger & 0 & \gamma  & 0 & 0& \\
 0 & 0 & v\pi & 0 & 0 & 0 & 0 & 0& \\
 0 & 0 & 0 & 0 & 0 & v\pi^\dagger& 0 & 0& \\
 0 & 0 & \gamma  & 0 & v\pi & 0 & \gamma  & 0 & \\
 0 & 0 & 0 & 0 & 0 & \gamma  & 0 & v\pi^\dagger&  \\
 0 & 0 & 0 & 0 & 0 & 0 & v\pi & 0& \\
  &  &  &  &  &  &  & & \ddots \\
\end{array}
\right).
\end{equation}
Here $\pi=\xi p_x+i p_y$, with valley index $\xi=\pm$. The Fermi velocity is $v=\frac{\sqrt{3}}{2}at_0$, where $t_0$ is the intralayer hopping parameter and $a$ is the lattice constant. $\gamma=t_\perp$ is the interlayer hopping.
The Schr\"odinger equation becomes $H|\psi_r^\pm\rangle=\varepsilon_r^{\pm}|\psi_r^\pm\rangle$ where the $2N$ eigenfunctions are
\begin{equation}
    |\psi_r^\pm\rangle=\begin{pmatrix}
    \varphi_{A_1}\\
    \varphi_{B_1}\\
    \varphi_{A_2}\\
    \varphi_{B_2}\\
    \vdots\\
    \varphi_{A_N}\\
    \varphi_{B_N}
    \end{pmatrix}
\end{equation}
and the eigenenergies are
\begin{equation}
\label{eq:eigenenergies}
    \varepsilon_r^{\pm}=\gamma\cos\bigg(\frac{r\pi}{N+1}\bigg)\pm\sqrt{(vp)^2+\gamma^2\cos\bigg(\frac{r\pi}{N+1}\bigg)^2}
\end{equation}
for $r=1...N$. Let us focus on $N$ even for now, in which case there are $2N$ quadratic bands, of which $N$ bands, $|\psi_r^-\rangle$, are at low energies (gapless). For odd $N$ there is also a Dirac cone and we will avoid the complications coming from that situation. 
The low energy bands are
\begin{equation}
\varepsilon_r^-(p)=
\left\{\begin{array}{ll}\frac{p^2}{2m_r}  & \textrm{if } \cos\bigg(\frac{r\pi}{N+1}\bigg)<0\\ 
-\frac{p^2}{2m_r}  & \textrm{if } \cos\bigg(\frac{r\pi}{N+1}\bigg)>0
\end{array} \right. 
\end{equation} 
where
\begin{equation}
    m_r=\frac{\bigg|\gamma\cos\bigg(\frac{r\pi}{N+1}\bigg)\bigg|}{v^2}=2m^*\bigg|\cos\bigg(\frac{r\pi}{N+1}\bigg)\bigg|
\end{equation}
where $m^*=\gamma/2v^2$. So the bands come in pairs with the same effective mass $m_r$, the bands related by $r+r'=N+1$ are such pairs. Let us call them conjugate bands.

\subsection{Low energy effective theory}
In the low energy limit $\varepsilon_r\ll pv$ we can also write down a low-energy effective Hamiltonian. The Schr\"odinger equation of the full Hamiltonian is
\begin{equation}
    v\pi^\dagger \varphi_{B_{2n-1}}=\varepsilon \varphi_{A_{2n-1}}
\end{equation}
\begin{equation}
    \gamma(\varphi_{A_{2n-2}}+\varphi_{A_{2n}})+v\pi \varphi_{A_{2n-1}}=\varepsilon \varphi_{B_{2n-1}}
\end{equation}
\begin{equation}
    v\pi \varphi_{A_{2n}}=\varepsilon \varphi_{B_{2n}}
\end{equation}
\begin{equation}
    \gamma(\varphi_{B_{2n-1}}+\varphi_{B_{2n+1}})+v\pi^\dagger \varphi_{B_{2n}}=\varepsilon \varphi_{A_{2n}}
\end{equation}

We can now eliminate $\varphi_{A_{2n}}$ and $\varphi_{B_{2n-1}}$ from these equations and use $\varepsilon\ll \pi v$. We can then write these equations as the Schr\"odinger equation for the simpler effective Hamiltonian

\begin{equation}
    H_{\textrm{eff}}=h+h^\dagger
\end{equation}
where
\begin{equation}
    h=-\frac{1}{2m^*}
    \begin{pmatrix}
     0& 0 & 0&0 &0 &0  &\dots \\
    \pi^2& 0& 0  &0 &0 &0 &\dots\\
     0& 0 &0 & 0&0 &0 &\dots \\
    -\pi^2& 0&\pi^2 &0 & 0&0&\dots \\
    0& 0& 0& 0&0 & 0&\dots\\
    \pi^2& 0&-\pi^2 &0 &\pi^2  &0&\dots\\
    \vdots&\vdots&\vdots&\vdots&\vdots&\vdots&\ddots
    \end{pmatrix}
\end{equation}
To solve this, we note that $\varphi_{A_0}=\varphi_{A_{N+1}}=0$ and $\varphi_{B_0}=\varphi_{B_{N+1}}=0$ so we try the ansatz
\begin{equation}
    \varphi_{A_n}=A\sin \bigg(\frac{n r\pi}{N+1}\bigg),\qquad \varphi_{B_n}=B\sin \bigg(\frac{n r\pi}{N+1}\bigg)
\end{equation}
This ansatz works and the reduced (Bloch) wavefunction is then (for the $K$ valley)
\begin{equation}
    |\psi_r(\vec{p})\rangle=\begin{pmatrix}
    \varphi_{A_1}\\
    \varphi_{B_1}\\
    \varphi_{A_2}\\
    \varphi_{B_2}\\
    \vdots\\
    \varphi_{A_N}\\
    \varphi_{B_N}
    \end{pmatrix}=\sqrt{\frac{2}{N+1}}\begin{pmatrix}
    e^{-2i\theta_p}\sin r\pi/(N+1)\\
    \sin 2r\pi/(N+1)\\
    e^{-2i\theta_p}\sin 3r\pi/(N+1)\\
    \vdots
    \end{pmatrix}
\end{equation}
with eigenvalues $\varepsilon_r^-(\vec{p})$.
We can easily see that for $N=2$ we obtain the same results as previously for BLG.

\subsection{Matrix elements}
We define the matrix elements $M_{rr'}(\vec{p},\vec{p'})$ as
\begin{align}
    M_{rr'}(\vec{p},\vec{p'})&\equiv\langle \psi_{r'}(\vec p')|\psi_{r}(\vec p)\rangle\\
    &=\frac{2}{N+1}\bigg[e^{-2i(\theta_p-\theta_{p'})}\sum_{\textrm{n odd}}\sin\bigg(\frac{nr\pi}{N+1}\bigg)\sin\bigg(\frac{nr'\pi}{N+1}\bigg)\\
    &+\sum_{\textrm{n even}}\sin\bigg(\frac{nr\pi}{N+1}\bigg)\sin\bigg(\frac{nr'\pi}{N+1}\bigg)\bigg]
\end{align}
where $1\leq n\leq N$. Using the trigonometric identities
\begin{equation}
\sum_{\textrm{n odd}}\sin\bigg(\frac{nr\pi}{N+1}\bigg)\sin\bigg(\frac{nr'\pi}{N+1}\bigg)=
\left\{\begin{array}{ll}\frac{N+1}{4}  & \textrm{if }r=r'\\ 
\frac{N+1}{4}  & \textrm{if }r+r'=N+1\\ 
0  & \textrm{otherwise}
\end{array} \right. 
\end{equation} 
and

\begin{equation}
\sum_{\textrm{n even}}\sin\bigg(\frac{nr\pi}{N+1}\bigg)\sin\bigg(\frac{nr'\pi}{N+1}\bigg)=
\left\{\begin{array}{ll}\frac{N+1}{4}  & \textrm{if }r=r'\\ 
-\frac{N+1}{4}  & \textrm{if }r+r'=N+1\\ 
0  & \textrm{otherwise}
\end{array}, \right. 
\end{equation} 
we obtain

\begin{equation}
 M_{rr'}(\vec{p},\vec{p'})=
\left\{\begin{array}{ll}\frac{1}{2}(1+e^{-2i(\theta_p-\theta_{p'})})  & \textrm{if }r=r'\\ 
\frac{1}{2}(-1+e^{-2i(\theta_p-\theta_{p'})})  & \textrm{if }r+r'=N+1\\ 
0  & \textrm{otherwise}
\end{array} \right. 
\end{equation} 
We will find it useful to introduce a more appropriate notation for the even $N$ case, namely $(r,N+1-r)\to(R,\sigma)$ where $R=r\ \textrm{mod}(\frac{N}{2}+1)=1...N/2$ and $\sigma=+,-$, where
\begin{equation}
\sigma=
\left\{\begin{array}{ll}+  & \textrm{if } \cos\bigg(\frac{r\pi}{N+1}\bigg)<0\qquad \textrm{i.e.}\ r> N/2\\ 
-  & \textrm{if } \cos\bigg(\frac{r\pi}{N+1}\bigg)>0\qquad \textrm{i.e.}\ r\leq N/2
\end{array} \right. 
\end{equation} 
In this notation we have paired up conjugate bands and hence
\begin{equation}
    \varepsilon_{R\sigma}(p)=\sigma\frac{p^2}{2m_R}
\end{equation}
and
\begin{equation}
    M_{R\sigma,R'\sigma'}(\vec p,\vec p')=\delta_{RR'}\frac{1}{2}(\sigma\sigma'+e^{-2i(\theta_p-\theta_{p'})})
\end{equation}
Now just make a slight redefinition of our wavefunctions
\begin{equation}
    \psi_{R\sigma}(\vec{p})=\sigma \psi_r(\vec{p})
\end{equation}
and with this additional sign
\begin{equation}
\label{eq:M}
    M_{R\sigma,R'\sigma'}(\vec p,\vec p')=\delta_{RR'}\frac{1}{2}(1+\sigma\sigma'e^{-2i(\theta_p-\theta_{p'})})
\end{equation}
so we just have $N/2$ copies of the BLG matrix elements, labelled by $R$ and where we denote particle-hole index by $\sigma$. 

We have $\pi=p_x+ip_y$ in the $K$ band and $\pi=-p_x+ip_y$ in the $K'$ band. So to treat the $K'$ band we need to replace $\pi_{K'}=-\pi_K^\dagger$. Since only $\pi^2$ appears in the Hamiltonian, we obtain the $K'$ wavefunctions from the $K$ wavefunctions by simple complex conjugation. So the matrix elements will also be complex conjugates of each other. However, we have a freedom to choose the overall phase of our wavefunctions, and this allows us to redefine our wavefunctions to cancel off this complex conjugation and we end up with the same matrix elements as for the $K$ valley. Therefore, the valley degeneracy can be taken into account simply by including a factor of $N_f=2\times2$ for the number of fermion species in the calculation (the additional factor of $2$ comes from spin degeneracy).

The charge density operator can be derived in the same manner as in Ref.~\cite{ourPRB}, and we obtain the result
\begin{equation}
\label{eq:rho}
    \rho(\mathbf{q})=\sum_{f}\sum_{R R'}\sum_{\sigma \sigma'}\int \frac{d^2 \mathbf{k}}{(2\pi)^2} c^\dagger_{R\sigma f}(\mathbf{k})c_{R'\sigma' f}(\mathbf{k+q})M_{R\sigma,R'\sigma'}(\vec{k},\vec{k+q}),
\end{equation}
where $c^\dagger_{R\sigma f} (c_{R'\sigma'f})$ is the creation (annihilation) operator of an electron.
The result shows that the Coulomb vertex will not allow transitions between bands with different masses and different flavors due to the explicit form of $M_{R\sigma,R'\sigma'}(\vec{k},\vec{k+q})$ in Eq.~\eqref{eq:M}.

\section{RPA screening calculation}
\label{sec:Screen}
In this section, we calculate the screened Coulomb potential in the random phase approximation (RPA). We use the explicit form of the density operator \eqref{eq:rho} and consider the RPA diagram Fig. \ref{fig:Feynman}. 
\begin{figure}
    \centering
    \includegraphics[width=0.5\linewidth]{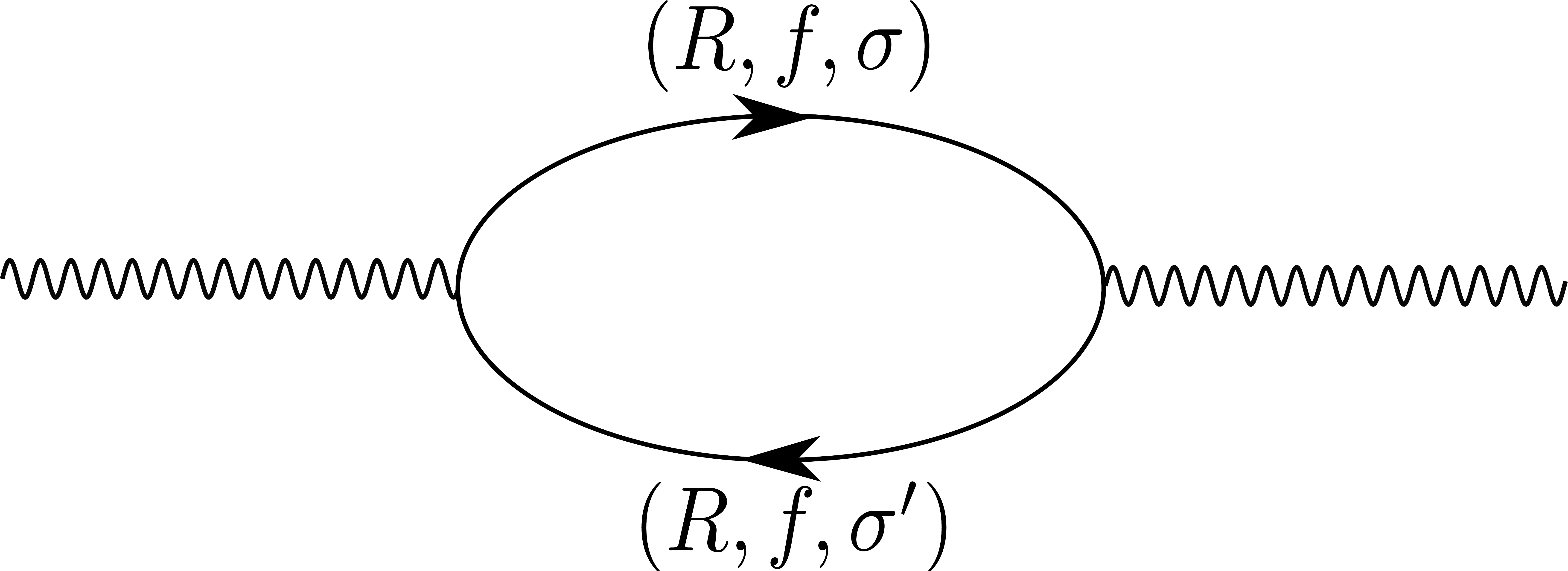}
    \caption{Feynman diagram for polarization. Note that due to the form of the vertex, the two electrons have the same value of $R$ and $f$. }
    \label{fig:Feynman}
\end{figure}
One can calculate the RPA polarizability and obtain
\begin{equation}
    \Pi^0(\Vec{q},0)=-N_f\sum_{\sigma,\sigma'}\sum_{R,R'}\int \frac{d^2k}{(2\pi)^2}\frac{f^{R'\sigma'}(\vec{k}+\Vec{q})-f^{R\sigma}(\vec{k})}{\varepsilon_{R'\sigma'}(\vec{k}+\Vec{q})-\varepsilon_{R\sigma}(\vec{k})}|M_{R\sigma,R'\sigma'}(\vec{k},\vec{k+q})|^2
\end{equation}
and using the $\delta_{RR'}$ in the matrix elements
\begin{equation}
    \Pi^0(\Vec{q},0)=-N_f\sum_{R}\sum_{\sigma,\sigma'}\int \frac{d^2k}{(2\pi)^2}\frac{f^{R\sigma'}(\vec{k}+\Vec{q})-f^{R\sigma}(\vec{k})}{\varepsilon_{R\sigma'}(\vec{k}+\Vec{q})-\varepsilon_{R\sigma}(\vec{k})}|M_{R\sigma,R\sigma'}(\vec{k},\vec{k+q})|^2
\end{equation}
But now for each $R$, the calculation is identical with the BLG case, so in the limit $\beta\mu\ll 1$
\begin{equation}
    \Pi^0(\Vec{q},0)=\sum_{R}\Pi^0_R(\Vec{q},0)=\frac{N_f}{2\pi}\sum_{R}m_R=\frac{N_fm^*}{2\pi}\sum_{R=1}^{N/2}2\cos\bigg(\frac{\pi R}{N+1}\bigg)=\frac{N_fm^*}{2\pi}\bigg(\frac{1}{\sin(\frac{\pi}{2(N+1)})}-1\bigg)
\end{equation}
\section{Details of quantum Boltzmann equation}
\label{sec:QBE}
In this section we follow Ref.~\cite{ourPRB}. In the low-energy bandstructure of multilayer graphene with an even number $N$ of layers, there are $N$ quadratic bands, which we label by $r=(R,\sigma)$, where $R=1,2,\cdots N/2$ and $\sigma=\pm$. The band energy is
\begin{equation}
    \varepsilon_{R\sigma}(p)=\sigma\frac{p^2}{2m_R}
\end{equation}
The band mass is $m_R=2m^* |\textrm{cos}((R\pi)/(N+1))|$, where $m^*=0.033m_e$. The equilibrium distribution of the electrons in band $r=(R,\sigma)$ is given by the Fermi distribution
\begin{equation}
f_r(\vec{p})=f^0(\epsilon_r(\vec{p}))=\frac{1}{1+e^{\beta(\epsilon_r(\vec{p})-\mu)}}.
\end{equation}
We write the deviation from the equilibrium distribution as
\begin{equation}
f_r(\vec{p})=f^0(\epsilon_r(\vec{p}))+f^0(\epsilon_r(\vec{p}))[1-f^0(\epsilon_r(\vec{p}))]h_r(\vec{p})
\end{equation} 
and expand the Boltzmann equation up to first order in $h_r(\vec{p})$. The Boltzmann equation is now a set of $N$ equations
\begin{equation}
\label{eq:QBE}
\frac{2\pi\sigma\beta}{m_r}f^0_r(\mathbf{p})[1-f^0_r(\mathbf{p})]\bigg(
e\mathbf{E}\cdot\mathbf{p}  
-\frac{1}{T}\nabla T\cdot\mathbf{p}(\epsilon_r(\mathbf{p})-\mu)\bigg) =I_r^{\textrm{tot}}[h_{r_i} (\vec{k}_i)](\vec{p})
\end{equation}
The LHS of the QBE includes the driving force due to the electric field $\vec E$ and the thermal gradient $\vec \nabla T$. The collision integral is 
\begin{align}
I_r^{\textrm{tot}}[h_{r_i} (\vec{k}_i)](\vec{p})=I_{r,\textrm{Coul}}^{(1)}[h_{r_i} (\vec{k}_i)](\vec{p})
&-\frac{1}{\tau_{ep,r}}f_r^0(p)[1-f_r^0(p)]h_r(\vec{p})
\label{eq:C_I}
\end{align}
The second term on the RHS of Eq.~\eqref{eq:C_I} is the contribution to the collision integral coming from electron-phonon collisions, for which the scattering rate is
    \begin{equation}
        \tau_{ep,r}^{-1}=\frac{D^2m_r k_BT}{2\rho\hbar^3c^2},
    \end{equation}
where $D$ is the deformation potential, $\rho$ is the mass density of multilayer graphene and $c$ is the speed of sound. Let us define the corresponding dimensionless number
    \begin{equation}
        \alpha_{ep}=\beta\tau_{ep}^{-1}=\frac{D^2m^*}{2\rho\hbar^3c^2}
    \end{equation}
The first term on the RHS of Eq.~\eqref{eq:C_I} is the linearized collision integral for scattering between electrons which is
\begin{align}
\label{eq:CoI1}
I_{r,\textrm{Coul}}^{(1)}[h_{r_i} (\vec{k}_i)](\vec{p})&=-(2\pi)\sum_{r_1r_2r_3}\int \frac{\diff^2\vec{k}}{(2\pi)^2} \frac{\diff^2\vec{q}}{(2\pi)^2}\delta(\epsilon_{r}(\vec{p})+\epsilon_{r_1}(\vec{k})-\epsilon_{r_2}(\vec{p+q})-\epsilon_{r_3}(\vec{k-q}))\nonumber\\
&\times\bigg[N_f|T_{rr_1r_3r_2}(\vec{p},\vec{k},\vec{q})|^2-T_{rr_1r_3r_2}(\vec{p},\vec{k},\vec{q})T_{rr_1r_2r_3}^*(\vec{p},\vec{k},\vec{k-p-q})\bigg] \\
&\times \bigg[[1-f_r^0(\vec{p})][1-f_{r_1}^0(\vec{k})]f_{r_2}^0(\vec{p+q})f_{r_3}^0(\vec{k-q})\bigg]\nonumber\\
&\times\bigg[-h_{r}(\vec{p})-h_{r_1}(\vec{k})+h_{r_2}(\vec{p+q})+h_{r_3}(\vec{k-q})\bigg]\nonumber
\end{align}
The matrix elements in \eqref{eq:CoI1} are
\begin{equation}
	T_{r_1r_2r_3r_4}(\Vec{k},\Vec{k'},\Vec{q})=V(-\Vec{q})M_{r_1r_4}(\Vec{k}+\Vec{q},\Vec{k})M_{r_2r_3}(\Vec{k'-q},\Vec{k'})
\end{equation}
with
\begin{equation}
    M_{r,r'}(\vec p,\vec p')=\delta_{RR'}\frac{1}{2}(1+\sigma\sigma'e^{-2i(\theta_p-\theta_{p'})})
\end{equation}
and with screened Coulomb potential
\begin{equation}
    V(\vec q)=\frac{2\pi}{N_fm^*}\bigg(\frac{1}{\sin(\frac{\pi}{2(N+1)})}-1\bigg)^{-1}
\end{equation}
The equations for the charge current and heat current are
		\begin{equation}
	  \mathbf{J}=  N_fe\sum_{r}\int \frac{d^2k}{(2\pi)^2} \frac{\sigma \mk}{m_r} f_{r}(\mk),
	\end{equation}

		\begin{equation}
	  \mathbf{J}^{Q}=  N_f\sum_{r}\int \frac{d^2k}{(2\pi)^2} \frac{\sigma \mk}{m_r}(\epsilon_{r}(\mk)-\mu) f_{r}(\mk).
	\end{equation}
In the case where we only have an applied electric field $\vec E$, the suggested ansatz to solve the QBE \eqref{eq:QBE} is \cite{ourPRL,ourPRB,Lux2012}
	\begin{equation}
	\label{eq:as1}
	h_r(\mathbf{p})=\beta\frac{e\mathbf{E}}{m^*}\cdot\mathbf{p}\chi_r(p).
	\end{equation}
We expand \eqref{eq:as1} in terms of basis functions
	\begin{equation}	
		\chi_r(k)=\beta\sum_n a_n g_n(r,k)
	\end{equation}
such that the $a_n$ are dimensionless. Here the basis functions are taken to be
	\begin{equation}
		g_n(r,k)=\delta_{r=1},\delta_{r=2},...\delta_{r=N},\delta_{r=1}K,\delta_{r=2}K,...\delta_{r=N}K,...
	\end{equation}
where $K=\sqrt{\beta/m}k$ is the dimensionless momentum. For all powers $n>2$ we multiply by an exponential factor so the basis function is $K^ne^{-K/2}$. We expand in up to $4N$ basis functions. Increasing the number of basis function changes the results only marginally. 
We use the fact that this must be valid for all $\vec E$, sum over $r$, multiply separately by $\vec{\hat{p}}g_m(r,p)$ and integrate over $\vec p$. This yields an equation that can be summarized in matrix form as
\begin{equation}
	M_{mn}a_n=F_m
	\label{eq:matrixEq}
\end{equation}
where we defined the dimensionless matrices
\begin{equation}
	M_{mn}=\beta \bigg(\frac{\beta}{m}\bigg)^{3/2}\sum_r\int \frac{d^2\mathbf{p}}{(2\pi)^2}g_m(r,p) I_r\bigg[\bigg\{\vec{\hat{p}}\cdot\mathbf{k}_ig_n(r_i,k_i)\bigg\}\bigg](\mathbf{p})
\end{equation}
and the dimensionless vector
\begin{equation}
	F_m= \beta\bigg(\frac{\beta}{m}\bigg)^{1/2}\sum_r\int \frac{d^2\mathbf{p}}{(2\pi)^2} \frac{\sigma p}{m_r} f^0_r(p)[1-f^0_r(p)]g_m(r,p)
\end{equation}
\eqref{eq:matrixEq} can be inverted to yield $\vec a$. The charge current is
	
	\begin{align}
	\mathbf{J}&=N_f\sum_r \frac{e}{m_r}\int \frac{d^2\mathbf{p}}{(2\pi)^2}\sigma \mathbf{p}f_r(\mathbf{p})\\&=\beta N_f\sum_r \int \frac{d^2\mathbf{p}}{(2\pi)^2}\sigma \mathbf{p}f^0_r(\mathbf{p})[1-f^0_r(\mathbf{p})]\frac{e^2\mathbf{E}}{m^*m_r}\cdot \mathbf{p}\chi_r(p).\nonumber
	\end{align}
	The DC conductivity is read off as 
	\begin{equation}
	\sigma_{xx}=\beta N_f\sum_r \int \frac{d^2\mathbf{p}}{(2\pi)^2}\sigma f^0_r(p)[1-f^0_r(p)]\frac{e^2p_x^2}{m_rm^{*}}\chi_r(p)=\frac{N_f e^2}{2\hbar}\vec G\cdot M^{-1} \vec F,
	\end{equation}
where we have exceptionally restored $\hbar$ and where the dimensionless vector
\begin{equation}
G_m=\beta\bigg(\frac{\beta}{m}\bigg)\sum_r\int \frac{d^2\mathbf{p}}{(2\pi)^2} \frac{\sigma p^2}{m_r} f^0_r(p)[1-f^0_r(p)]g_m(r,p).
\end{equation}
The thermal conductivity and thermopower can be calculated completely analogously.

\section{ Multi-fluid model}\label{sec:mFM}
We can now derive a multi-fluid model. We assume that the electrons/holes in band $(R,+/-)$ have mean velocity $\vec u^{(R,+/-)}$. The corresponding ansatz for the perturbation of the distribution function is $h_{r} (\vec{k})=\beta \vec{k}\cdot \vec u^r$. We obtain the fluid equations by multiplying the QBE \eqref{eq:QBE} by $\vec p$ and integrating over $\vec p$. We then divide by the number density $n^r$ to obtain the coupled set of equations
\begin{equation}
    m_r\partial_t\vec{u}^r=-\sum_{r'}\frac{m_r}{\tau_{rr'}}(\vec{u}^r-\vec{u}^{r'})-\frac{m_r\vec{u}^r}{\tau_{ep,r}}+\sigma e\vec{E}-\Lambda^r k_B\nabla T.
    \label{eq:fluid1}
\end{equation}
Remember that we have defined $e<0$ as the electron charge. We define $\Lambda^r$ through the integrals
\begin{equation}
\label{eq:Lambda}
    \Lambda^{(R,+)}=\frac{\int \frac{d^2 \vec{p}}{(2\pi)^2} \frac{\beta p^2}{m_r}\beta(\epsilon_r(p)-\mu)f^0_r(\mathbf{p})[1-f^0_r(\mathbf{p})]}{\int \frac{d^2 \vec{p}}{(2\pi)^2} f^0_r(\mathbf{p})}
\end{equation}
\begin{equation}
    \Lambda^{(R,-)}=\frac{\int \frac{d^2 \vec{p}}{(2\pi)^2} \frac{\beta p^2}{m_r}\beta(-\epsilon_r(p)+\mu)f^0_r(\mathbf{p})[1-f^0_r(\mathbf{p})]}{\int \frac{d^2 \vec{p}}{(2\pi)^2} (1-f^0_r(\mathbf{p}))}.
\end{equation}
Note that $\Lambda^r=\Lambda^{(R,\sigma)}$ only depends on $\sigma$, not $r$ itself, since the species mass $m_r$ drops out when we de-dimensionalize. So this is in fact exactly the same expression as in BLG. These integrals are in fact the entropy per particle 
\begin{equation} 
\Lambda^{(R,+)}=-\frac{\int \frac{d^{2}\vec p}{(2 \pi)^{2}}\left[\left(1-f^0_r(\mathbf{p})\right) \ln \left(1-f^0_r(\mathbf{p})\right)+f^0_r(\mathbf{p}) \ln f^0_r(\mathbf{p}) \right]}{\int \frac{d^{2} \vec p}{(2 \pi)^{2}} f^0_r(\mathbf{p})} ,
\end{equation}

\begin{equation} 
\Lambda^{(R,-)}=-\frac{\int \frac{d^{2}\vec p}{(2 \pi)^{2}}\left[\left(1-f^0_r(\mathbf{p})\right) \ln \left(1-f^0_r(\mathbf{p})\right)+f^0_r(\mathbf{p}) \ln f^0_r(\mathbf{p}) \right]}{\int \frac{d^{2} \vec p}{(2 \pi)^{2}} [1-f^0_r(\mathbf{p})]} ,
\end{equation}The number density of species $r$ is 
\begin{equation}
    n^r=\frac{N_fm_r}{2\pi\beta}\ln(1+e^{\sigma\beta\mu})
\end{equation}
where $\sigma=+/-$ depending on whether we are dealing with a particle or a hole band. In order to obtain the scattering time between species $r$ and $r'$ due to Coulomb interactions, we need to solve the equation
\begin{equation}
\int \frac{d^2 \vec{p}}{(2\pi)^2} \vec{p}I_{r,\textrm{Coul}}^{(1)}\bigg[h_{r_i} (\vec{k}_{r_i})=\beta \vec{k}_{r_i}\cdot \vec{u}^{r_i}\bigg](\vec{p})=-\sum_{r'}\frac{n^rm_r}{\tau_{rr'}}(\vec{u}^r-\vec{u}^{r'}).
\label{eq:Coul_drag}
\end{equation} 
Instead of explicitly computing this collision integral for all $\beta\mu$ and all $r$, a reasonable first guess (that needs to be checked against the QBE results) would be
\begin{equation}
    \tau_{rr'}^{-1}=\frac{n^{r'}}{\sum_{r''}n^{r''}}\bigg(\frac{m_{r}}{\sum_{r''}m_{r''}}\bigg)^{-1/2}\tau_{ee}^{-1}
\end{equation}
such that we only need to evaluate $\tau_{ee}$. To see where this guess comes from, remember that basic kinetic theory yields $\tau_{rr'}^{-1}\sim n^{r'}\Sigma \langle v_r\rangle$ ($\Sigma$ is the collision cross-section) and $\langle v_r\rangle\sim\sqrt{k_BT/m_r}$. We then plug the steady-state solution of the fluid equations \eqref{eq:fluid1} into the formula for the electrical current
\begin{equation}
    \vec{J}=e\sum_r\sigma n^r\vec{u}^r
    \label{eq:J_current}
\end{equation}
and heat current
\begin{equation}
    \vec{Q}= k_BT\sum_r\Lambda^rn^r\vec{u}^r.
    \label{eq:Q_current}
\end{equation}
We can change this into a 1d problem in the absence of a magnetic field. To this end consider the steady-state form of equation \eqref{eq:fluid1}
\begin{equation}
    0=-\sum_{r'}\frac{m_r}{\tau_{rr'}}(\vec{u}^r-\vec{u}^{r'})-\frac{m_r\vec{u}^r}{\tau_{ep,r}}+\sigma e\vec{E}-\Lambda^r k_B\nabla T
    \label{eq:steady_state}
\end{equation}
We can turn this into a matrix equation by defining the scattering rate between species as
\begin{equation}
    \Gamma_{rr'}=\delta_{rr'}\bigg(\sum_{r''}\tau_{rr''}^{-1}+\tau_{ep,r}^{-1}\bigg)-\tau_{rr'}^{-1}
\end{equation}
such that \eqref{eq:steady_state} becomes
\begin{equation}
    m_r\sum_{r'}\Gamma_{rr'}u^{r'}=\sigma eE-\Lambda^r k_B\nabla T.
\end{equation}{} 
The solution is obtained by taking the matrix inverse
\begin{equation}
    u^{r}=\sum_{r'}\frac{1}{m_r}\Gamma_{rr'}^{-1}(\sigma' eE-\Lambda^{r'} k_B\nabla T)
    \label{eq:soln_for_u}
\end{equation}{}
Plugging \eqref{eq:soln_for_u} into the electrical current \eqref{eq:J_current} one finds
\begin{equation}
    \vec{J}=e\sum_r\sigma n^r\sum_{r'}\frac{1}{m_r}\Gamma_{rr'}^{-1}(\sigma'eE-\Lambda^{r'} k_B\nabla T)
\end{equation}
and similarly for the heat current \eqref{eq:Q_current} one obtains
\begin{equation}
    \vec{Q}= k_BT\sum_r\Lambda^rn^r\sum_{r'}\frac{1}{m_r}\Gamma_{rr'}^{-1}(\sigma'eE-\Lambda^{r'} k_B\nabla T).
\end{equation}
Finally, this leads to the expressions for the electrical conductivity
\begin{equation}
    \sigma_{xx}=e^2\sum_{rr'}\frac{n^r}{m_r}\sigma\sigma'\Gamma_{rr'}^{-1}=N_fe^2\sum_{rr'} \sigma\sigma'\tilde n^r\alpha_{rr'}^{-1}\equiv \frac{N_fe^2}{2} \tilde\sigma_{xx}
\end{equation}{}
the thermopower
\begin{equation}
    \Theta_{xx}=N_fek_B\sum_{rr'}\frac{n^r}{m_r}\Gamma_{rr'}^{-1}\Lambda^{r'}\sigma=N_fek_B\sum_{rr'}\sigma\tilde n^r\alpha_{rr'}^{-1}\Lambda^{r'}\equiv \frac{N_fek_B}{2} \tilde\Theta_{xx}
\end{equation}{}
and the thermal conductivity (modulo the usual caveat about the open-circuit thermal conductivity)
\begin{equation}
    K_{xx}=k_B^2T\sum_{rr'}\frac{n^r}{m_r}\Lambda^r\Gamma_{rr'}^{-1}\Lambda^{r'}=N_fk_B^2T\sum_{rr'}\tilde n^r\Lambda^r\alpha_{rr'}^{-1}\Lambda^{r'}\equiv \frac{N_fk_B^2T}{2} \tilde K_{xx}
\end{equation}{}
where we have de-dimensionalized by defining
\begin{equation}
    \tilde n^r=\beta\frac{n^r}{N_fm_r}=\frac{\ln(1+e^{\sigma\beta\mu})}{2\pi}
\end{equation}
and
\begin{equation}
    \alpha_{rr'}=\beta\Gamma_{rr'}.
\end{equation}{}
In analogy with BLG we define dimensionless tildered quantities, which for conciseness we reproduce here again
\begin{equation}
    \tilde\sigma_{xx}=2\sum_{rr'} \sigma\sigma'\tilde n^r\alpha_{rr'}^{-1}
\end{equation}{}
\begin{equation}
    \tilde\Theta_{xx}=2\sum_{rr'}\sigma\tilde n^r\alpha_{rr'}^{-1}\Lambda^{r'}
\end{equation}{}
\begin{equation}
    \tilde K_{xx}=2\sum_{rr'}\tilde n^r\Lambda^r\alpha_{rr'}^{-1}\Lambda^{r'}.
\end{equation}{}

\section{Supplementary figures}\label{sec:sup_figs}

\begin{figure}[H]
    \centering
    \includegraphics[width=0.5\linewidth]{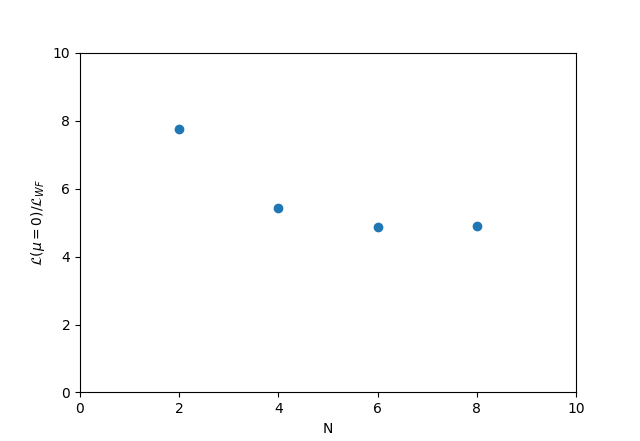}
    \caption{Results for $\mathcal L(\mu=0)/\mathcal L_{WF}$, where $\mathcal L=\sigma/\kappa T$ is the Lorenz number and $\mathcal L_{WF}=\pi^2/3(k_B^2/e^2)$ is the Wiedemann-Franz result. We present the QBE results for different even values of $N$. We have set $\alpha_{ep}=0.1/N$ as in our previous work on BLG.}
    \label{fig:WF_N}
\end{figure}

\begin{figure}[H]
    \centering
    \includegraphics[width=0.5\linewidth]{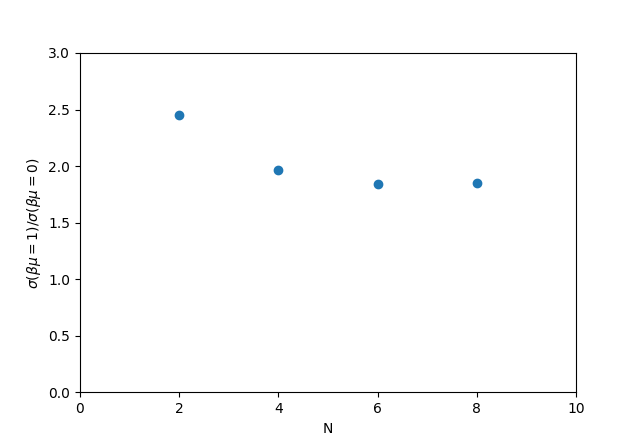}
    \caption{Results for $\tilde \sigma(\beta\mu=1)/\tilde \sigma(\beta\mu=0)$ from the QBE calculation for different even values of $N$. We have set $\alpha_{ep}=0.1/N$ as in our previous work on BLG.}
    \label{fig:sigma_curvature_N}
\end{figure}

\end{widetext}

\bibliography{bib}

\end{document}